# Unstable and elusive superconductors

Yakov Kopelevich, Robson R. da Silva, and Bruno C. Camargo

Instituto de Física "Gleb Wataghin", Universidade Estadual de Campinas, Unicamp 13083-859, Campinas, São Paulo, Brasil

## ABSTRACT

We briefly review earlier and report original experimental results in the context of metastable or possible superconducting materials. We show that applied electric field induces conducting state in Copper Chloride (CuCl) whose characteristics resemble behavior of sliding charge-density-wave(s) (CDW). We discuss whether the sliding CDW or collective transport of similar ordered charge phase(s) may account for the problem of "high-temperature superconductivity" observed in this and other materials, including Cadmium Sulfide (CdS), metal-ammonia solutions, polymers, amorphous carbon and tungsten oxides. We also discuss a local superconductivity that occurs at the surface of graphite and amorphous carbon under deposition of foreign atoms/molecules.

## 1. History and state of the art.

Since the discovery in 1911 by Heike Kamerligh Onnes of the superconductivity (SC) in mercury (Hg) with the transition temperature $T_c = 4.2$ K, an empirical approach in searching for new superconductors with higher $T_c$ is still most successful one.

In recent times, the most spectacular achievement of such approach was the discovery of high-temperature superconducting (HTS) cuprates [1], allowing to use a practical liquid nitrogen (T = 77 K) instead of expensive liquid helium (T = 4.2 K) to reach the superconducting state. So far, the highest $T_c$ = 135 K at ambient pressure was achieved for copper mercury oxide $HgBa_2Ca_2Cu_3O_9$ [2] and $T_c$ was raised up to 164 K under pressure of 30 GPa [3].

The discovery of HTS left little doubts among researchers that the superconductivity at room temperature RTS (better to say at normal leaving conditions) is achievable.

There were published many review articles on the progress in superconducting materials, and routes to RTS, see e. g. [4 -7].

*In the present contribution we briefly describe some earlier reports on potential high-$T_c$ superconductors and present our experimental results that shed more light on this and related long-standing problems. We shall only concentrate on the results duplicated by at least one independent experimental group.*

Looking back on the history of superconducting materials, one sees that discovery of novel superconductors is often associated with initial findings of traces of unstable (transient) superconductivity due to presence of a small amount of a novel superconducting phase embedded in the parent material.

### 1.1. Cuprate high-$T_c$ superconductors.

In the beginning of 1987, C. W. Chu and collaborators have observed superconducting activities up to ~ 96 K in mixed-phase La-Ba-Cu-O samples [8] that have been lost in one day because of the sample degradation. Later on, the famous superconducting $LaBa_2Cu_3O_y$ (La-123) phase with $T_c \geq 90$ K was identified.

Various experimental groups reported a metastable, i. e. vanishing with time, superconducting response in Y-Ba-Cu-O samples even at room temperature.

For instance, Munger and Smith measured reverse ac Josephson effect in polycrystalline $YBa_2Cu_3O_{7-\delta}$ up to 300 K [9] and they have concluded that the RTS is related to the sample inhomogeneity. Remarkably, after 25 years, reports on metastable RTS in cuprates continue to appear in the literature. In Ref. [10] a transient RTS superconductivity has been triggered by light in $YBa_2Cu_3O_{6+\delta}$, attributed to the light-induced enhancement of the Josephson inter-layer coupling. Before that, conclusions on the enhanced coupling between neighboring Cu (2) - O (2) bi-layers caused by photo-assisted oxygen ordering [11] was made [12]. Hence, thermally unstable structures may be behind of transient RTS in cuprates.

A strong enhancement of $T_c$, as compared to the equilibrium bulk value, resulting from the surface adsorption of nitrogen, oxygen, helium has been also demonstrated [13, 14]. Then, desorption of foreign atoms/molecules during the thermal cycling would naturally account for the $T_c$ reduction, and hence, the metastable superconductivity.

## 1. 2. Elusive high-temperature superconductors.

### 1.2. A. Copper Chloride and Cadmium Sulfide.

Before the cuprate era, experiments performed on Copper Chloride (CuCl) [15-22] and Cadmium Sulfide (CdS) [23-25] were a major hope related to high-temperature superconductivity.

In 1978, Brandt et al. [15] reported a transition to the almost ideal diamagnetic state in CuCl below $T_d \sim 170$ K, accompanied by the resistivity drop of several orders of magnitude, when the samples under a hydrostatic pressure of ~ 5 kbar were rapidly (~ 20 K/min) cooled/heated. The "super-diamagnetic" state appears to be metastable such that $T_d$ has decreased gradually by a factor of ~ 3 after several thermal cycles in the temperature range 4.2 K < T < 350 K. However, at constant temperature, the anomalous state was stable for hours. The results revealed a jumpy behavior of ac susceptibility between "zero" and "super-diamagnetic" state values. Similar results were obtained in the temperature interval $\Delta T \sim 20$ K around T ~ 240 K at slightly different conditions but again only under a rapid sample warming [16, 17]. It was proposed [16] that under applied pressure and rapid temperature variation, the reaction $2CuCl \rightarrow Cu + CuCl_2$ leads to formation of Cu-CuCl metal-insulator metastable interfaces responsible for the transient HTS. The assumption that Cu solely is responsible for the observed diamagnetism requires the copper resistivity far smaller than known values.

While the pressure-induced insulator-metal transition (IMT) in CuCl has been widely explored, a similar IMT induced by applied voltage (V) [18, 19] is less known. Gentile [18] showed that the electric-field (~ 3 kV/cm) driven IMT in CuCl is accompanied by the formation of Cu filaments, and Divakar et al. [19] demonstrated that both applied pressure and electric field drive the IMT such that the critical (threshold) electric field $E_{th}$ decreases with the pressure increase. Besides, it has been observed that the insulating state re-appears removing the applied voltage. The recovery time depends on various factors, and it could be as long as 3 days when the conducting state kept for 48 hours [19].

The almost ideal diamagnetism [20], as well as the low resistance state (LRS) [21] below ~ 240 K were also observed in CuCl films grown on high purity Si(111) substrates. DC magnetization M(H) measured with the SQUID magnetometer was found to be paramagnetic or diamagnetic when the magnetic field (H) applied parallel or perpendicular to the CuCl/Si interface, respectively [20]. Such anisotropy implies a two-dimensional (2D) character of the phenomenon related to either the sample surface or CuCl/Si interface [20]. The authors [20] optically detected islands of the size as large as ~ 0.1 mm that break with time on smaller domains and form a hexagonal array during tens of days. The observed domain structure was not affected by a small variation of magnetic field or temperature but it was sensitive to mechanical deformations.

We recall that Auger electron spectroscopy and low-energy ion scattering studies [22] proved that CuCl/Si(111) interface is thermodynamically unstable, viz. Cu forms the outermost layer of CuCl film implying the Si-Cl bonding at the interface. Hence, it is possible that similar to bulk CuCl, formation of Cu/CuCl interfaces is responsible for the observed superconducting-like behavior.

Similar to CuCl, pressure-quenched CdS containing Cl impurities (0.6-0.8 wt %) has been found to demonstrate superconducting-like behavior at temperatures well above 77 K [23, 24]. The samples were prepared under applied pressure P > 40 kbar that was released at rates > $10^6$ bar/s. The quenched "as-received" samples demonstrated 5 orders of magnitude lower resistance and superconducting-like magnetization hysteresis loops M(H) with nearly perfect Meissner portion. These properties depend crucially on the Cl contents as well as pressuring, thermal and magnetic history. The experiments revealed that the metastable diamagnetism decays as the electrical conductivity. In Ref. [25] x-ray diffraction and photo-acoustic measurements have been made on CdS as a function of chlorine doping. The results revealed the Cl-induced cubic to hexagonal transformation accompanied by the presence of soft lattice modes over a critical range of Cl impurities where the superconducting-like state takes place. Furthermore, authors of Ref. [25] developed a theoretical model suggesting the occurrence of Fröhlich-type superconductivity [26] in CdS. In the Fröhlich theory there is no electron-electron pairing but superconductivity may result from a freely sliding charge density waves (CDW). When sliding CDW travel around closed loops, strongly diamagnetic state accompanied by the resistance drop should take place.

### 1.2. B. Polypropylene polymers.

It is instructive to compare results obtained on CuCl and that reported for oxidized atatic polypropylene (OAPP) polymers [27-29].

Thick (0.3 - 100 μm) insulating polypropylene films were deposited on Cu or In substrate and after some thermal cycling the apparent zero-resistance state below T ~ 300 K has appeared [27, 28]. Similar to CuCl, OAPP demonstrate a strong metastable diamagnetism. The reported results revealed that the initial sate of OAPP is weakly diamagnetic. After the field cycling in the interval 0 < B < 0.2 T, the samples were found to be paramagnetic, ferromagnetic or strongly diamagnetic, resembling the behavior of superconductors in a partially penetrated Meissner state [27, 29]. The experiments repeatedly showed spontaneous transformations between ferromagnetic- (FM) and SC-like states at room temperature. The abrupt decrease in the resistance measured for OAPP [28] upon increasing the current is similar to that observed for CuCl/Si films [21].

### 1.2. C. Metal-ammonia solutions.

The phenomenon of current/voltage - driven highly conducting state measured in CuCl and OAPP, has been also reported for much older potential high-temperature superconductor, viz. decomposing metal(M)-ammonia ($M-NH_3$) solution where M = Li, Na [30]. Thus, measurements of current-voltage (I-V) characteristics at T = 240 K performed on $Na-NH_3$ [31] revealed that above a certain critical voltage ($V_{th}$), a sudden increase of the current takes place signaling the transition towards a highly conducting state. The observed behavior resembles the transition between pinned and nearly free sliding charge density wave (CDW) state. This similarity led Arendt [31] to propose the Fröhlich-type superconductivity occurrence in $Na-NH_3$. In the metallic state the resistance of $Na-NH_3$ decreases exponentially $R(T) \sim \exp(-aT_0/T)$ [32], with a ≈ 3, $T_0$ = 190 K suggesting the gap opening in the electronic spectrum, in contrast to gapless electronic spectrum in ordinary metals. To the best of our knowledge, gaped electron systems that demonstrate metallic-like behavior are only the systems with superconducting paring or charge (spin) density waves in sliding regime. Supporting either superconducting or CDW scenario "ring experiments" revealed a persistent electric current in $M-NH_3$ at T = 77 K [30].

### 1.2 D. Amorphous carbon.

Several decades ago, Antonowicz has measured Josephson-type oscillations at room temperature in Al-AC-Al sandwiches (AC = amorphous carbon) [33, 34]. The resistance drop by ~ two orders of magnitude has been induced by applied electric field ~ 100 kV/cm, in a close analogy with the experiments realized on CuCl, OAPP and $M-NH_3$ (see above). Once generated, the LRS has persisted for several days before an initial high-resistance state (~ 5·10$^8$ Ω) re-established. Observed maximum current oscillations vs. magnetic field $I_{max}(H)$ as well as Shapiro-like steps in I-V characteristics measured in the presence of ~ 10 GHz microwave field [34], led the author to conclude on a possible room-temperature superconductivity associated with conducting filaments in the low resistance phase.

### 1.2. E. Tungsten oxides.

In context of the Fröhlich superconductivity, we would like to mention a resistance drop observed in amorphous $WO_x$ (a-$WO_x$) films, grown on glass substrates, below a temperature $T^*(I, f)$ that is current (I) - and frequency (f) - dependent [35]. The LRS was triggered by higher current or/and frequency, such that for I = 1 mA and f = 1 kHz, the resistance drop in the temperature interval 2K < T < $T^*$ ~ 20 K was as large as two orders of magnitude. In this regime $R(T) = R_0 + R_1 \exp(-T_0/T)$, with $T_0$ = 24 K. Overall the results suggest that LRS in a-$WO_x$ originates from moving (sliding) many-body interacting electron system such as Wigner crystal [35].

It would be interesting to explore more the interplay between structural disorder, CDW and superconductivity in tungsten oxides. Indeed, doping of $WO_3$ leads to the superconductivity below ~ 4 K [36] or CDW state at $T_{CDW}$ ~ 150 K [37]. Results reported in Refs. [38-40] suggest the occurrence of superconductivity with $T_c$ = 90 K in Na-doped $WO_3$. STM measurements [39, 40] showed that possible superconductivity occurs within randomly arranged metastable Na-rich islands of 20-150 nm size at the sample surface or Na/$WO_3$ interfaces, implying its 2D character.

### 1.2. F. Surfaces and interfaces.

In general, adsorbed at the surface atoms/molecules can be responsible for the superconductivity localized only within surface layers of the atomic thickness, indeed. Several theoretical studies were devoted to this problem in the past. In the model proposed by Ginzburg [41], conducting electrons (holes) interact with adsorbed atoms at the sample surface leading to an effective attraction between the surface carriers. According to BCS theory, $k_B T_c$ = ℏωexp[-1/N(0)V], where N(0) denotes the density of states at the Fermi level and V is the mean matrix element of the interaction energy corresponding to attraction. Taking a typical difference between energy levels of adsorbed atoms ℏω ~ 1 eV, and considering the characteristic value for N(0)V ≈ 0.4...0.5, one gets $T_c$ ~ $10^3$ K.

Agranovich et al. [42] showed that because of the resonance character of scattering from adsorbed atoms, electron-electron attraction at the surface would exceed that in the bulk, resulting in the high-temperature surface superconductivity.

A recent proposal was to use adsorbed non-magnetic molecules to create negative-U centers, and hence "islands" of electron singlet pairs, on the surface of materials that have Dirac-type electronic spectrum [43]. It has been also shown that a combination of attractive interaction and disorder-induced enhancement of density of states N(0) trigger local SC in Dirac semimetals [44].

Motivated by experiments that suggest the occurrence of local room temperature superconductivity in graphite [45], Gonzálves et al. [46] showed that topological disorder in graphite or graphene (one single layer of graphite) enhances density of states and triggers the p-wave superconductivity when electron-electron repulsive interaction is strong enough.

It has been demonstrated both experimentally [47 - 50] and theoretically [51] that sulfur (S) atoms adsorbed on graphene/graphite trigger the local superconductivity, and $T_c(0) \geq$ 35 K [47, 48, 50] has been measured. In Refs. [52 - 54] researchers explored the possibility of superconductivity in graphene triggered by adsorbed alkali metals. In angle-resolved photoemission spectroscopy (ARPES) experiments performed on graphene decorated by Ca atoms, the superconducting gap $\Delta_s$ ~ 10 meV has been detected [52]. Also, the superconductivity with $T_c$ = 1.5 K has been reported for Ca-decorated graphene in Ref. [53]. Magnetization measurements performed on K-doped few-layer-graphene revealed a localized superconductivity with $T_c$ = 4.5 K, which is about 30 times higher than that observed for K-intercalated graphite [54].

We conclude this section citing the famous sentence by Wolfgang Pauli: "God made the bulk; surfaces were invented by the devil" reflecting that fact that surface properties strongly depend on an external environment, making the experimental work quite challenging. Nevertheless, the above examples tell us that sample surfaces and interfaces deserve more attention in searching for superconductors with higher $T_c$. One of the most spectacular recent achievements in this direction is the enhanced interface superconductivity in one-unit-cell FeSe films grown on $SrTiO_3$ [55, 56], possibly occurring above T = 100 K [57].

In what follows we present our experimental results that shed more light on the origin of some phenomena described above.

## 2. Our recent experiments.

### 2. 1. Electric-field-driven insulator-metal transition in CuCl.

Motivated by findings [18, 19], we explored the effect of applied electric field on transport properties of CuCl. Measurements of the resistance R(T, B) and I-V characteristics were performed on compressed with P = 1 GPa, CuCl powder of the grain size ~ 40 μm (Sigma Aldrich) of 99.995% purity in the temperature interval 2 K < T < 300 K and magnetic field up to B = 9 T using Janis He-4 cryostat. The powder were compressed into cylindrical pellets of 5mm diameter and 1mm thickness. In two- or four-probe resistance measurements we used silver (Ag) paste electrodes placed on the sample surface in vertical or planar geometry.

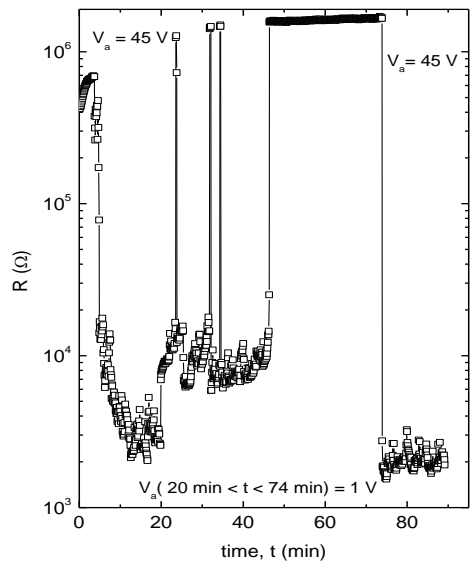

Fig. 1. Electric-field-driven transition between high- and low-resistance states in CuCl and its intermittent time dependence.

Figure 1 presents time variation of the two-probe resistance R(t) recorded at T = 300 K under applied voltage $V_a$ for one of our CuCl samples. As can be seen from Fig. 1, under $V_a$ = 45 V, R(t) varies slowly during the first minutes and then suddenly drops by ~ 300 times. At $t_1$ = 20 min, $V_a$ has been reduced to 1 V, resulting in the intermittent time relaxation between low and high resistance states for $t_1 < t < t_2$ = 74 min. At t = $t_2$, $V_a$ has been again increased to 45 V inducing the sharp transition towards low-resistance state. We did not see such transition in samples prepared from greenish oxidized CuCl powder.

The electric breakdown due to segregation of free Cu in the form of filaments of a few microns diameter has been proposed [18]. However, oscillations between two resistance values (see Fig. 1) cannot be easily understood assuming just formation of Cu "wires". One can also argue against a simple Zener model for dielectric breakdown. Indeed, taking the gap energy in CuCl, $E_g$ = 3.4 eV [58] and a lattice constant a = 5.41Å [59] one estimates the threshold field for dielectric breakdown $E_{th}$ ~ 70 MV/cm which is much higher than the applied in experiments field E ~ 450 V/cm. No insulator-metal was observed at temperatures T < 200 K within the same time window and $V_a \leq$ 100 V. In samples made of ~ 400 μm size particles, that possess a smaller surface area, the resistance drop at the transition is very modest (factor ~ 4) suggesting that the conductivity is governed by surface of the grains. Besides, we measured similar voltage-driven transition with two electrodes placed on the same sample surface, i. e. in the "planar geometry".

Noting that the bi-stability effect similar to that shown in Fig. 1, was also observed in other systems that exhibit voltage-controlled insulator-metal [60] or CDW [61] transitions.

It is found that the low-resistance state in CuCl is characterized by non-linear I-V characteristics, I ~ $V^\alpha$ with α = 3/2 (Fig. 2), the behavior known for CDW systems in a vicinity of the electric-field-driven depinning transition [62]. The right inset in Fig. 2 provides evidence for the "metallic" (dR/dT > 0) origin of the conducting state. We did not detect the metallic state in four-probe measurements suggesting that metallic phase occurs within narrow channels.

One can speculate that applied at T = 300 K electric field E > $E_{th}$ delocalizes preexisting CDW leading to the highly conducting state. Alternatively, the low-resistance state can be associated with the appearance of Cu filaments followed by nucleation of CDW at Cu/CuCl interfaces. Then, CDW freely sliding around closed loops can account for enhancement of both diamagnetism and electrical conductivity [15-17] without invoking the high-temperature superconductivity in a usual sense [61].

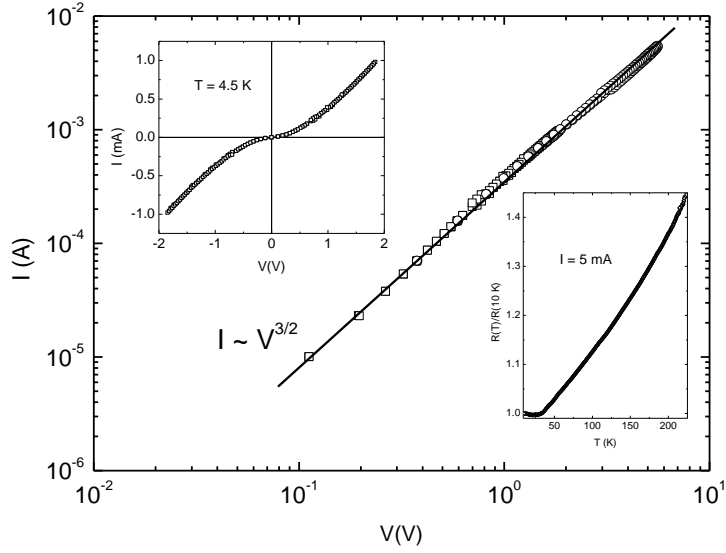

Fig. 2. Two-probe I-V characteristics measured at T = 4.5 K (□) and T = 4.15 K (o). The solid line corresponds to the power law I ~ $V^{3/2}$. The left inset shows I-V measured at T = 4.5 K for both positive and negative applied voltage. The right inset presents a temperature variation of the reduced resistance R(T)/R(10K) in the electric-field-induced metallic state, measured at constant applied current I = 5 mA.

## 1.2. Enhanced transient superconductivity in reduced $WO_{3-x}$.

Inspired by a possible occurrence of high-temperature surface superconductivity in Na-doped $WO_3$ [38-40] as well as reports on twin wall superconductivity in reduced $WO_{3-x}$ crystals [63], we studied crystalline $WO_{3-x}$ films obtained by means of the pulsed laser deposition [64].

The results presented in Figures 3-5 were obtained for as-received $WO_{3-x}$ film [64] of size 6 mm x 6 mm x 400 nm grown at T = 400 °C and oxygen pressure ~ $10^{-2}$ mbar. The film possesses a granular structure with the grain size of a few nanometers or less that form 100-200 nm agglomerates. We performed resistivity measurements in the van der Pauw geometry with four point-like silver paste contacts using Janis He-4 cryostat. The measured resistivity $\rho(T_c)$ ~ 11 mΩcm (Fig. 3) suggests x ~ 0.4 ($WO_{2.96}$) [65].

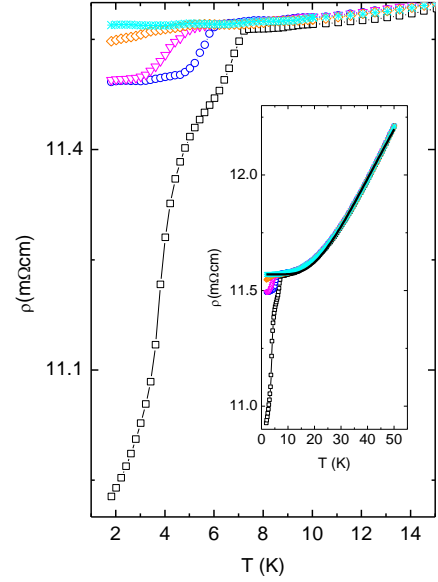

Fig. 3. Resistively measured superconducting transition measured for as received $WO_{3-x}$ film for applied perpendicular to the films main surface magnetic fields H = 0 (□), 0.5 kOe (o), 1 kOe (∇), 1.9 kOe (◊), 3.99 kOe (*). The inset shows the same ρ(T) data up to T = 50 K; solid line corresponds to the equation $\rho(T) = \rho_0 + \rho_1 \exp(-T_0/T)$, where $\rho_0$ = 11.57 mΩcm, $\rho_1$ = 3.1 mΩcm, $T_0$ = 83 K.

Figure 3 illustrates the resistivity drop measured below the field dependent transition temperature $T_c(H)$ that becomes undetectable for $H \geq 3.99$ kOe. At zero applied field, a two-step transition is evident suggesting the superconductivity occurring between weakly coupled superconducting grains [66], and only one-step transition takes place for $H \geq 500$ Oe implying the field-induced suppression of the intergrain Josephson coupling. Figure 4 presents $d\rho/dT$ vs. T obtained for $H = 0$ and $H = 200$ Oe, where maxima in $d\rho/dT$ vs T mark two distinct transition temperatures $T_c(H) > T_{cJ}(H)$ such that $T_{cJ}(H)$ is much more sensitive to the applied magnetic field. $T_c(H)$ and $T_{cJ}(H)$ obtained for various fields are given in Figure 5.

upper critical field $H_{c2}(T)$ of isolated (decoupled) superconducting grains [67], and Eq.(2) describes the critical field $H_{cJ}(T)$ for the intergrain Josephson medium [68, 69]. Because $H_{cJ} \sim \sqrt{j_{cJ}}$, the linear dependence $H_{cJ}(T) \sim [1 - T/T_{cJ}(0)]$ implies $j_{cJ}(T) \sim [1- T/T_{cJ}(0)]^2$ for the critical current density of superconductor-normal metal - superconductor (SNS) type junctions [70]. Using equations $H_{cJ} = (12\mu_0\Phi_0 j_{cJ}/d)^{1/2} = [24\pi E_J(T_c)/d^3]^{1/2}$, $E_J(T_{cJ}) \sim k_B T_{cJ} = \hbar j_{cJ} d^2/2e$ [68, 69], together with the measured $T_{cJ}(0) = 3.8$ K and $H_{cJ}(0) = 400$ Oe (Fig. 5), we estimate the grain size $d \leq 10$ nm and $j_{cJ} \sim 10^5$ A/cm$^2$. The estimated grain size practically coincides with that measured with SEM [64], and calculated $j_{cJ}$ agrees nicely with characteristic values of $j_{cJ} \sim 10^4$-$10^5$ A/cm$^2$ measured in HTS thin film weak links [71].

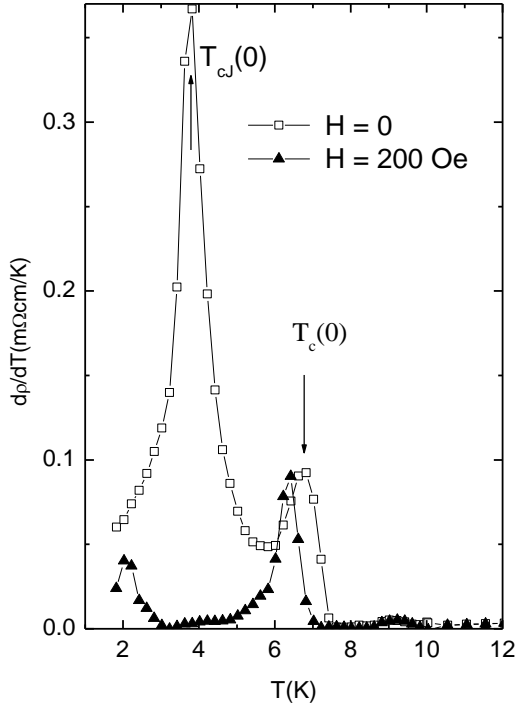

Fig. 4. Two peaks in $d\rho/dT$ vs. T correspond to superconducting transition temperatures related to isolated grains $T_c(H)$ and intergrain Josephson coupling $T_{cJ}(H)$.

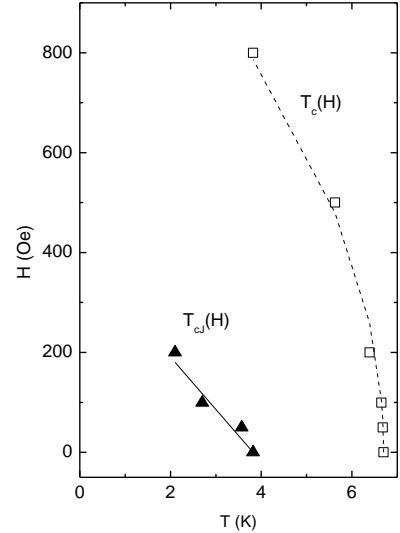

Fig. 5. Magnetic field - temperature (H-T) phase diagram demonstrating the upper critical field $T_c(H)$ boundary of decoupled superconducting grains and the inter-grain Josephson coupling transition temperature $T_{cJ}(H)$, as determined in Fig. 4. Dashed and solid lines correspond to Eqs. (1) and (2), respectively with the fitting parameters $T_c(0) = 6.7$ K, $H_0 = 1.2$ kOe, $T_{cJ}(0) = 3.8$ K, and $H_{cJ}(0) = 400$ Oe.

One describes very well the data by the equations:

$$H_{c2}(T) = H_0[1 - T/T_c(0)]^{1/2} \quad (1)$$

$$H_{cJ}(T) = H_{cJ}(0)[1 - T/T_{cJ}(0)] \quad (2)$$

From Eqs. (1) and (2) we get $T_c(0) = 6.7 \pm 0.1$ K, $H_0 = 1.2$ kOe, $T_{cJ}(0) = 3.8 \pm 0.1$ K, and $H_{cJ}(0) = 400$ Oe. Eq. (1) corresponds to the

We kept the film at ambient conditions for 12 months and then we measured it again. As Figure 6 illustrates, the superconductivity is gone and $\rho(T)$ shows a shallow maximum at $T \sim 50$ K. The absolute value of the resistivity has increased from its initial value by a factor of $\sim 5$. Because metallic and superconducting behavior of WO$_{3-x}$ is due to electron doping originating from de-oxygenation, the film re-oxygenation in the time interval between two measurements would explain the resistivity increase and eventually the superconductivity lost. Thus, $WO_{3-x}$ provides us with an example of surface transient superconductivity with $T_c$ twice the value of that associated with twin walls [63].

characteristic of superconductors after 14 days keeping the sample at ambient conditions.

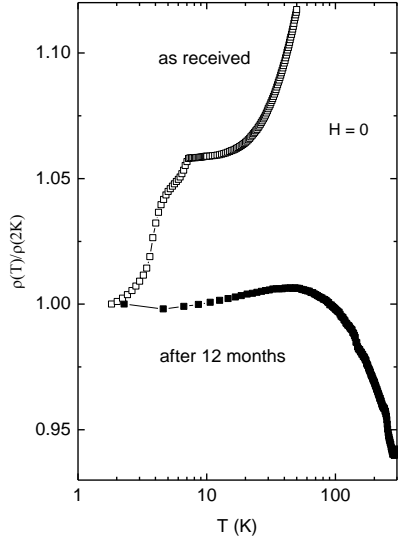

Fig. 6. Reduced resistivity obtained for as received (superconducting) $WO_{3-x}$ film and after keeping the film at ambient conditions for one year.

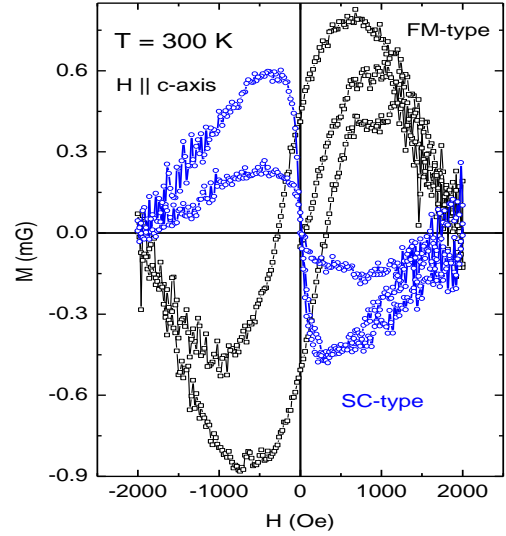

Fig. 7. (Color online)Ferromagnetic (FM) - and superconducting (SC) -type magnetization hysteresis loops obtained immediately and after two weeks of the heat treatment (see text), subtracting diamagnetic background magnetization $M = -\chi H$ with $\chi = 3.46 \cdot 10^{-2}$ mG/Oe and $\chi = 3.49 \cdot 10^{-2}$ mG/Oe, respectively.

## 2.3. Transient superconductivity in oxidized and sulfur-doped graphitic materials.

Temporal transformations between superconducting- and ferromagnetic-type states observed for both OAPP [27-29] and graphite [45] led us to explore in more details the oxygen effect on magnetic properties of graphite.

The results given in Figures 7-9 were obtained for highly oriented pyrolytic graphite (HOPG) sample from Union Carbide Co. subjected to the heat treatment at $T = 870$ K under low vacuum (~ 0.05 mbar) during 2 hours. The measurements were made using SQUID magnetometer (Quantum Design) with magnetic field oriented either parallel ($H \parallel c$) or perpendicular ($H \parallel$ planes) the hexagonal c-axis.

As Fig. 7 illustrates, the FM-type hysteresis loop M(H) measured immediately after the heat treatment, transformed to the hysteresis loop

Noting, the FM → SC transformation detected in $H \parallel c$ measurements (Fig. 7), did not affect the FM response measured for $H \parallel$ graphene planes (Fig. 8), suggesting that the superconductivity is localized within graphene planes or related to the sample surface. The results provide also evidence for the coexistence of the superconductivity and ferromagnetism.

The estimated superconducting volume fraction as small as ~ $10^{-3}$ % is consistent with the surface-related non-percolating SC.

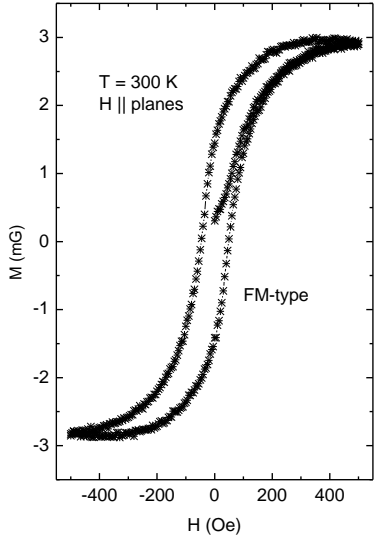

Fig. 8. M(H) hysteresis loop measured with H ∥ graphene planes for the SC sample after the background signal subtraction ($\chi = 5 \cdot 10^{-3}$ mG/Oe).

We have verified stability of the superconducting state (Fig. 7) to the additional sample heating. Figure 9 shows the superconducting-type hysteresis loop M(H) of Fig.7 together with the hysteresis loop measured immediately after heating the sample up to 350 K (in situ) (H ∥ c). As can be seen from Fig. 9, the heating results in a reduction of the hysteresis loop width. This observation testifies against of a possible artifact of measurements, at first place. Perhaps, most plausible explanation of the result would be the temperature-assisted oxygen adsorption-desorption process and/or oxygen migration to different defect sites at graphite surface.

According to [72], ferromagnetic in- and out-of-plane interactions transform to antiferromagnetic ones varying the oxygen contents at graphitic zig-zag edges. The results of Figs. 7 and 9 suggest that at some level of oxidation, the superconducting phase coexisting with the ferromagnetism is more favorable.

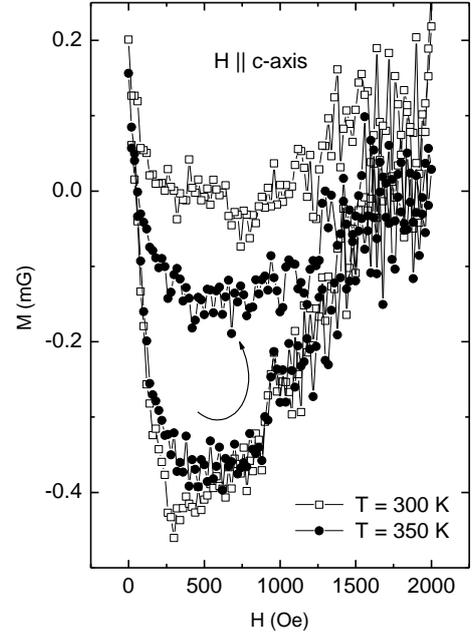

Fig. 9. Superconducting-like M(H) hysteresis loops obtained at T = 300 K (the same as in Fig. 7), and at T = 350 K measured immediately after heating the sample.

Similar to sulfur-doped graphite [49], the non-observability of FM component in the c-axis direction in transformed (SC) sample (Figs. 7, 9) implies the confining of the FM moment within graphene planes when the superconductivity is established. This may occur when the superconducting pairing breaks a time-reversal symmetry leading to the interaction between SC-induced and pre-existing ferromagnetic moments [49]. Supporting this scenario, time-reversal symmetry breaking p-wave [46] and d-wave superconductivity [51, 73, 74] has been proposed for doped graphene/graphite.

Overall results indicate that sulfur and oxygen play a similar role in the superconductivity occurrence in graphitic materials.

Here we would like to emphasize the following points related to the superconductivity in C-S composites [47-51]: (1) a broad variety of measured $T_c$, (2) granular character of the superconductivity, and (3) a graphitization role.

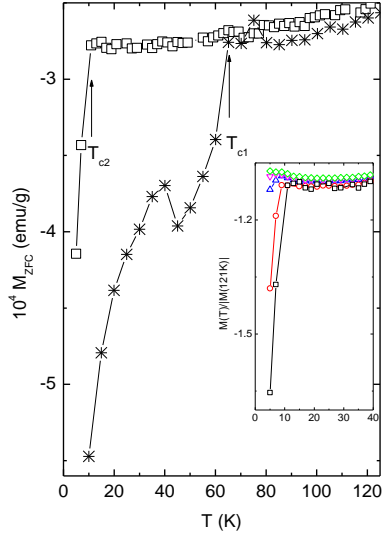

Fig. 10. (Color online) Zero-field-cooled magnetization, $M_{ZFC}(T)$ measured at H = 50 Oe for graphite-sulfur composite immediately after the sample synthesis (*) and after 11 days (□) with $T_{c1}$ = 65 K and $T_{c2}$ = 9 K, respectively. Inset demonstrates suppression of the superconducting transition by magnetic field: H = 50 Oe (□), 100 Oe (o), 300 Oe (Δ), 500 Oe (∇), 700 Oe (◊).

Figure 10 demonstrates two SC-type transitions measured for one of our sulfur-doped graphitic samples immediately after the sample synthesis ($T_{c1} \approx$ 65 K) and after keeping the sample at ambient conditions for 11 days ($T_{c2} \approx$ 9 K). Suppression of the transition by a weak magnetic field provides evidence for its superconducting nature, see inset in Fig. 10. In order to account for the transient superconductivity in C-S, we note (i) observation of two superconducting transitions at $T_c$ ~ 6.7 K and $T_c$ ~ 37 K in the same temperature run [48], and (ii) the enhancement of both $T_c$ and the superconducting volume fraction increasing the sulfur contents [50].

The superconductivity enhancement with the sulfur contents agrees with theoretical predictions of the $T_c$ rising with the sulfur doping level [51]. Then, an inhomogeneous decoration of graphite surface by sulfur atoms can account for two superconducting transitions reported in Ref. [48]. Rearrangement or desorption of sulfur atoms would also explain the results given in Fig. 10.

Figure 11 presents results of the resistance measurements performed on compressed with P = 1 GPa C-S composite with various measuring currents. As Fig. 11 illustrates, a sharp resistance drop at $T_c \approx$ 14 K takes place for I = 0.1 mA. For I = 1 mA, the apparent $T_c$ is shifted to T ~ 10 K, and the resistance drop becomes less pronounced. For I = 10 mA the transition cannot be seen down to the lowest measuring temperature T = 4.2 K. Such current-induced suppression of the superconductivity is known for granular superconductors [66, 75], and can be understood taking into account both an inter-grain Josephson vortex dynamics and current-driven Cooper pair dissociation inside the grains.

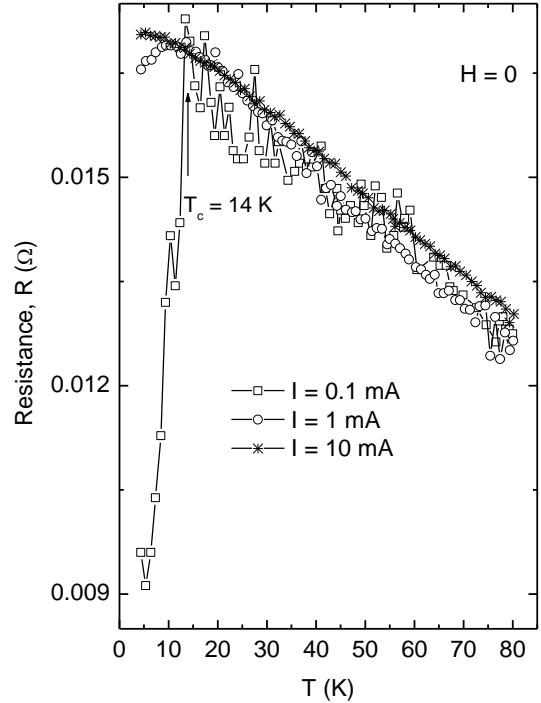

Fig. 11. Current-induced suppression of resistively measured superconducting-type transition in compressed sulfur-doped graphite powder.

The observation of SC-type behavior in sulfur-doped amorphous carbon (AC) [50] raised a question on the graphitization role in the superconductivity occurrence. AC is a strongly disordered material with $sp^3/sp^2$ carbon hybridization. It possesses submicron curved graphene layers with a mixed interlayer stacking. Experiments performed on nano-diamond [76] and amorphous carbon [77] showed that graphitization, i.e. $sp^3 \rightarrow sp^2$ transformation takes place at relatively low temperatures ~ 750 - 1000°C.

We prepared AC-S composites by mixing carbon glassy powder consisting of 10-40 μm size particles of 99.99 % purity and sulfur powder of 99.998% purity, both from Aldrich Chemical Company, Inc. in a mass ratio $m_{AC}:m_S = 1:1$ ($m_{AC} = m_S = 0.2$g). The mixture was pressed into pellets, sealed in evacuated (p = $10^{-2}$ mbar) quartz tube and then heated at 500 - 1200 °C for 24 h before cooling down to ambient temperature. We characterize the samples by means of x-ray diffraction (XRD) and scanning electron microscopy (SEM), and we measured sample magnetization M(H,T) using SQUID magnetometer from Quantum Design (MPMS5).

Figure 12 presents zero-field-cooled (ZFC) magnetization $M_{ZFC}(T, H)$ measured for virgin AC, and AC and AC-S heated at 800 °C. No signature for the superconductivity can be seen for virgin as well as heated AC down to T = 2 K. However, $M_{ZFC}(T, H)$ measured for sulfur-doped AC demonstrate behavior expected for the superconductor with $T_c \sim 3$ K (see also inset in Fig. 13). XRD patterns obtained for virgin and heated AC (inset in Fig. 12) revealed only broad features ("halo") characteristic of amorphous carbon.

Figure 13 presents XRD data for AC-S. In contrast to the heat treatment of AC without sulfur, a sharp diffraction peak at ~ 26.7 ° characteristic to Bernal graphite is clear. The inset in Fig. 13 shows the superconducting-type behavior of the AC-S sample below $T_c$ (H = 50 Oe) ≈ 3.25 K, suppressed in a weak magnetic field.

Figure 14 presents M(H) hysteresis loop measured at T = 2 K after cooling the sample from T = 300 K to the target temperature in a zero field. Shown in Fig. 14 magnetization hysteresis loop is characteristic of type-II superconductors with a strong vortex pinning; an abrupt reduction of the irreversible magnetization for |H| < 150 Oe (Fig. 14) resembles very much the behavior of type-II superconductors in the regime of thermo-magnetic flux jump instabilities [78].

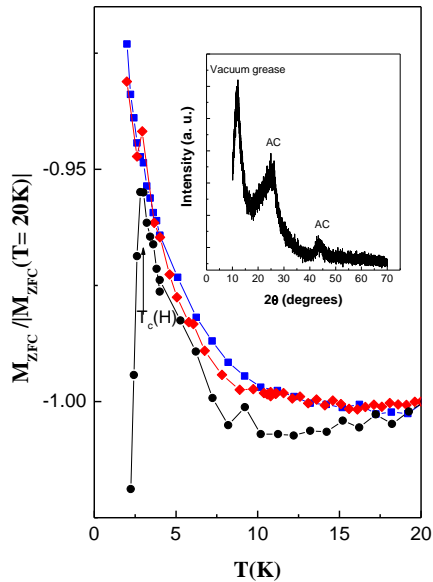

Fig. 12. (Color online) Normalized zero-field-cooled magnetization, measured with H = 100 Oe for virgin AC (■), heated at 800 °C AC (♦), and heated at 800 °C AC-S composite (●). Arrow marks superconducting transition temperature $T_c$(100 Oe) = 3 K. Inset gives the x-ray θ-2θ diffraction pattern measured for virgin and heated AC.

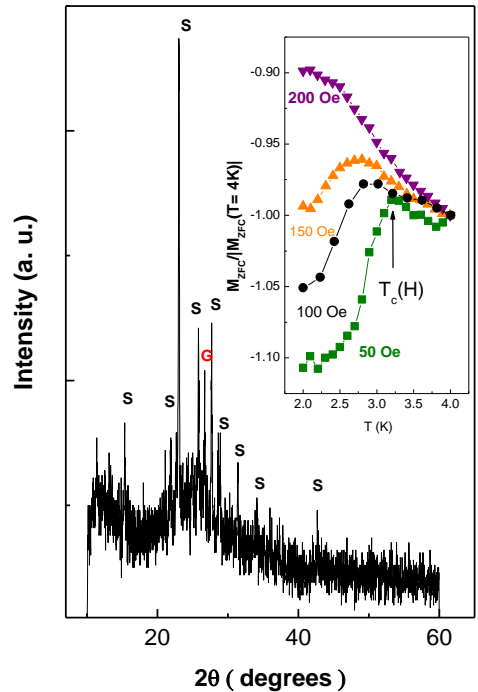

Fig. 13. (Color online) X-ray θ-2θ diffraction pattern measured for heated at 800 °C AC-S. Graphitization of AC is evident from the sharp peak at ~ 26.7 ° denoted as G. S-peaks correspond to the crystalline sulfur. Inset presents normalized ZFC magnetization $M_{ZFC}(T)/M_{ZFC}(4K)$ vs. temperature measured for various magnetic fields.

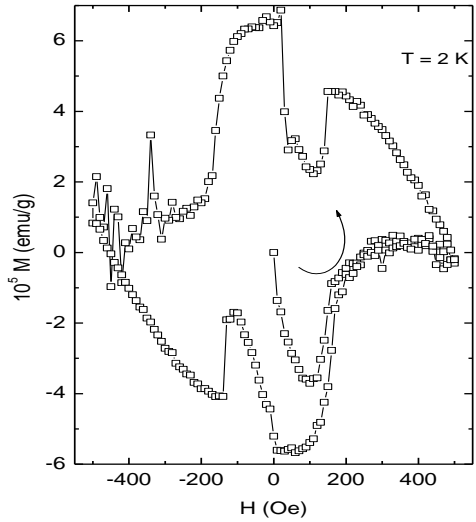

Fig.14. Magnetization hysteresis loop M(H) measured for the AC-S sample at T = 2 K < $T_c$ ~ 3 K (see Fig. 13) after subtraction of background magnetization M = - χH (χ = $2.15 \cdot 10^{-6}$ emu/g·Oe).

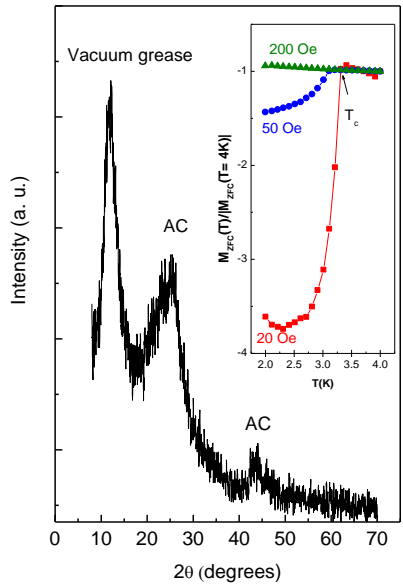

Fig. 15. (Color online) X-ray θ-2θ diffraction pattern measured for AC-S sample prepared at 500 °C. Inset presents normalized ZFC magnetization $M_{ZFC}(T)/M_{ZFC}(4K)$ vs. temperature measured for various magnetic fields demonstrating the occurrence of superconductivity with $T_c$ ~ 3.3 K.

Is the superconductivity in AC needs the graphitization (Fig. 13) ? The results given in Figure 15 for the AC-S sample synthesized at 500 °C give the answer to this question. As Fig. 15 illustrates, only "halo" due to amorphous structure appears in the XRD pattern. Nevertheless, the superconductivity takes place in this sample below $T_c$ ~ 3 K (inset in Fig. 15), similar to that measured for partially graphitized AC-S (Fig. 13).

Thus, the results given in Figs. 12 - 15 show that (1) sulfur stimulates the graphitization of amorphous carbon; (2) superconductivity in carbon-sulfur composites does not necessitate the presence of crystalline graphite; (3) similar to graphite-sulfur composites, $T_c$ in AC-S varies by order(s) of magnitude from sample to sample [50].

## 3. Concluding remarks and outlook.

(i) Our first remark is about the long-standing problem of possible high- temperature superconductivity in metal-ammonia solutions (M-NH$_3$) [30-32], oxidized atatic polypropylene (OAPP) polymers [27-29], CuCl [15-22], CdS [23-25], and amorphous carbon [33, 34].

Analysis of the literature data together with the results reported in section 2 indicates that Fröhlich-type (super)conductivity due to sliding CDW can account for both enhanced conductivity and diamagnetism measured in all these materials, without invoking the superconductivity in a usual sense.

Electric-field-induced low resistance state, non-linear I-V characteristics [I ~ exp(-1/V) or I ~$V^n$, n > 1], temporal fluctuations between low- and high-resistance states, periodic in magnetic field current (resistance) oscillations and Shapiro-like anomalies in I-V curves, all these are characteristic of CDW systems, see e. g. [61, 79, 80].

(ii) Amorphous $WO_x$ (x = 1.55) films, where the ac-current-induced transition to the low-resistance state was observed [35], is another potential Fröhlich-type superconducting system.

(iii) The transient superconductivity with $T_c$ ≈ 7 K measured in granular $WO_{2.6}$ films, twice of $T_c$ reported for $WO_{2.95}$ [63], suggests that the superconductivity with higher $T_c$ may occur in the samples with lower oxygen contents and smaller grain size.

(iv) Weak superconducting signals measured in partially oxidized and sulfur-doped graphite/carbon are presumably related to the sample surface. In a line with those results, Kawashima [81] has observed persistent current at room temperature during 50 days in ring-shaped compressed graphite flakes been in a contact with n-octane (saturated hydrocarbon).

Although not always measured, most of graphitic samples contain a large amount of hydrogen (~ $10^{15}$ H-atom/cm$^2$) concentrated near the surface [82]. Recently, superconductivity with $T_c$ ~ 190 K has been discovered in $SH_n$ (n > 2) hydrides under high pressure [83]. Hence, one should not exclude a priori the superconductivity associated with the formation of sulfur hydrides at the surface of graphitic materials. The authors of Ref. [83] also speculated on possible superconductivity in hydrocarbons. Atatic polypropylene ($C_3H_6$), discussed in the present paper, is a linear hydrocarbon where the methyl groups are placed randomly on both sites of the chain that can be easily oxidized. Thus, formation of C-H-O structures may be responsible for the superconducting-type magnetic response in both oxidized OAPP and graphite.

We thank Aline Rougier for providing us with $WO_{3-x}$ films. This work was supported by FAPESP, CNPq, CAPES, INCT NAMITEC, and AFOSR Grant FA9550-13-1-0056.